\documentclass[fleqn,usenatbib]{rasti}

\usepackage{newtxtext,newtxmath}
\usepackage[T1]{fontenc}

\DeclareRobustCommand{\VAN}[3]{#2}
\let\VANthebibliography\thebibliography
\def\thebibliography{\DeclareRobustCommand{\VAN}[3]{##3}\VANthebibliography}

\usepackage{graphicx}	% Including figure files
\usepackage{amsmath}	% Advanced maths commands
\usepackage{caption}

\title[When X-ray Features Fail to Identify Intrinsic Emitters]{When X-ray Features Fail to Identify Intrinsic Emitters: Label Noise and Luminosity Overlap in Machine Learning Classification of AGN and Star-forming Galaxies}

\author[J. Ding]{
Jaymin Ding$^{1}$\thanks{E-mail: jaymin@princeton.edu}
\\
$^{1}$Princeton University, Princeton, NJ 08544, USA\\
}

\date{Accepted XXX. Received YYY; in original form ZZZ}

\pubyear{\the\year{}}

\begin{document}
\label{firstpage}
\pagerange{\pageref{firstpage}--\pageref{lastpage}}
\maketitle

\begin{abstract}
\textrm{Previously}, we found that adding an X-ray flux feature to a Random Forest classifier of active galactic nuclei (AGN) and star-forming galaxies (SFGs) \textrm{coincided with a decrease in} classification accuracy from 97.51\% to 89.26\%, a counterintuitive result given the prevailing theory that X-rays are a reliable AGN diagnostic. \textrm{We investigate this} discrepancy through both astrophysical and machine learning lenses. On the astrophysical side, we show that the X-ray luminosities of AGN and SFGs in the sample substantially overlap across the canonical $10^{42}$ erg/s threshold, reflecting the moderate luminosity characteristic of an optically-selected sample and the contribution of high-mass X-ray binaries (HMXBs) to SFG emission. On the machine learning side, we argue that BPT-derived training labels \textrm{carry} instance-dependent label noise concentrated near the BPT demarcation line, where X-ray data would be most useful as a discriminator. We show that \textrm{objects misclassified} in 5-fold cross-validation cluster near the Kewley demarcation curve, with a median distance of \textrm{0.123} dex \textrm{versus} \textrm{0.743} dex for correctly classified objects (\textrm{Kolmogorov-Smirnov} $D = \textrm{0.745},\; p = \textrm{1.67}\times 10^{\textrm{-9}}$). \textrm{A controlled cross-validation decomposition shows that the previously reported decrease reflects predominantly sample selection\textrm{: on a fixed sample, adding the X-ray feature leaves class-resolved metrics unchanged, and its permutation importance is consistent with zero across resampled splits. The classifier therefore makes negligible use of a physically informative feature, consistent with training labels coupled to the optical features. We conclude that future classifiers operating on optically-selected samples should employ independent label sources or noise-robust training methods.}}
\end{abstract}

% Include between one and six keywords.

%%%%%%%%%%%%%%%%%%%%%%%%%%%%%%%%%%%%%%%%%%%%%%%%%%

%%%%%%%%%%%%%%%%% BODY OF PAPER %%%%%%%%%%%%%%%%%%

\section{Introduction}

Distinguishing star-forming galaxies (SFGs), galaxies whose luminosity is powered by stellar processes, from active galactic nuclei (AGN), galaxies whose luminosity is powered by accretion onto a supermassive black hole, is a central problem in extragalactic astronomy. Understanding how the relative populations of these two different types of galaxies form at different epochs of cosmic history is central to our understanding of how galaxies form, evolve, and quench their star formation over time. Accurate classification at scale is a prerequisite for population studies of galaxy evolution.

The gold standard for this classification, the Baldwin-Phillips-Terlevich (BPT) diagram, has served the field reliably since its introduction in 1981 \citep{1981PASP...93....5B}, using ratios of optical emission lines to separate the harder ionizing radiation of accreting black holes from the softer radiation of young, massive stars. However, its demarcation lines—most notably the \citet{2001ApJ...556..121K} theoretical maximum starburst line, hereafter referred to as the Kewley line—are statistical constructs derived from model grids and observational samples rather than true physical boundaries \citep{2013ApJ...774L..10K}. \textrm{The empirical demarcation of \citet{2003MNRAS.346.1055K}, which lies below the Kewley line and bounds the composite region where classification is most ambiguous, is also widely used. The BPT diagram for this sample is presented in \citet{2024PASP..136l4102D}}.

Previously, we introduced a Random Forest machine learning classifier trained on multi-wavelength photometric and spectroscopic data drawn from the Sloan Digital Sky Survey (SDSS) and the XMM-Newton Survey, achieving strong performance on the AGN/SFG classification task. However, one unexpected result was that when X-ray flux features were added to the model, its classification accuracy decreased from 97.51\% to 89.26\% rather than increasing. \textrm{This is a surprising outcome: it is counterintuitive that X-ray features fail to aid the classification of objects that are themselves intrinsic X-ray emitters. X-ray properties are known to be powerful discriminators for X-ray-bright source populations, for example, in the classification of galactic X-ray sources spanning active stars to accreting compact objects \citep{2024PASP..136e4201R}, which makes their lack of utility here in classifying accreting supermassive black holes all the more unexpected.} We previously noted that this result “could suggest that the X-ray properties of AGN and SFGs are more similar than previously thought” \citep{2024PASP..136l4102D}. In addition, peer review of our work raised the possibility that BPT-correlated training labels introduced a circularity distorting the apparent contribution of the X-ray features, and although the concern was addressed in further revisions, the underlying question was never definitively resolved. \textrm{We note at the outset that these two accuracies were measured on different samples: the full optically-selected sample and the smaller X-ray-detected subsample, a distinction we return to in Section~\ref{sec:ml}.}

The prevailing theory in X-ray astronomy holds that X-ray emission is one of the most reliable indicators of AGN activity, capable of identifying accreting black holes even when heavily obscured at optical wavelengths \citep{2018ARA&A..56..625H}. The physical basis for X-ray AGN selection rests on the accretion disk-corona model—where thermal UV emission from the accretion disk undergoes inverse Compton scattering off hot coronal electrons, producing a characteristic hard X-ray power-law spectrum \citep{1984ARA&A..22..471R,1991ApJ...380L..51H}—making AGN intrinsically X-ray luminous. If this theory is fully correct, then AGN and SFGs should have sufficiently distinct X-ray properties such that a machine learning model would benefit from incorporating the data. The \textrm{apparent decrease in} performance points to one of two explanations: either that the prevailing theory is incorrect or overstates the distinctiveness of AGN and SFG X-ray properties in the luminosity and redshift regime probed by this sample, or that something in the machine learning pipeline is preventing the model from extracting the discriminatory information the X-ray data contains. This paper evaluates both possibilities in turn, drawing on astrophysical considerations, instance-dependent label noise, and the boundary-concentration of misclassifications.

\section{Data and Methods}
\label{sec:methods}

\textrm{This analysis builds directly on the dataset and methodology of \citet{2024PASP..136l4102D}; we summarize the key details here for completeness. A parent sample was constructed from SDSS DR16 and cross-matched against the WISE, GALEX, and XMM-Newton source catalogues using TOPCAT \citep{2011ascl.soft01010T} within a 5 arcsecond matching radius. Objects without reliable detections, composite galaxies, and retired galaxies were excluded. The resulting sample contains 28,708 objects. The features used in both models are the optical spectral line ratios $\log\left(\text{[NII]}/\text{H}\alpha\right)$ and $\log\left(\text{[OIII]}/\text{H}\beta\right)$, optical photometric bands (u, g, r, i), near- and far-UV photometry, mid-infrared bands (W1, W2), spectroscopic redshift, and 0.2–12 keV X-ray flux. Objects without a confirmed XMM detection within the specified flux band were excluded from the X-ray model entirely, yielding 312 objects with X-ray measurements. \textrm{In \citet{2024PASP..136l4102D}, }both models were trained using an 80-20 train-test split with a Random Forest classifier \citep{2001MachL..45....5B} using default scikit-learn hyperparameters \citep{2011JMLR...12.2825P}, with the split repeated across multiple random states to confirm metric stability. \textrm{All models are evaluated with 5-fold stratified cross-validation, and quoted uncertainties are sample standard deviations across folds.}}

\textrm{Because the two accuracies reported in \citet{2024PASP..136l4102D} were measured on differently-sized samples, we present a controlled three-way cross-validation decomposition isolating the sample effect from the feature effect in Section~\ref{sec:ml}.}

\section{Astrophysical Considerations: Sample Luminosity Overlap}

Central to the theory is a luminosity threshold established by \citet{2018ARA&A..56..625H}: above approximately $10^{42}$ erg/s, X-ray emission can be attributed with high confidence to AGN activity, because ordinary stellar processes rarely produce emission at those luminosities in SFGs. However, this established threshold is not perfect. \citet{2012MNRAS.419.2095M} demonstrate that in SFGs, the collective X-ray emission from HMXBs can contribute substantially to the total observed X-ray luminosity and that such contribution scales with the star formation rate of the galaxy. In galaxies with high star formation rates, where large numbers of massive stars are being born and dying on relatively short timescales, the cumulative X-ray luminosity from these systems can approach or even exceed $10^{41}$ to $10^{42}$ erg/s, at which the boundary between AGN and SFG X-ray emissions becomes ambiguous. \textrm{More recent calibrations of the HMXB luminosity-SFR relation, including its metallicity dependence \citep{2016ApJ...825....7L,2016MNRAS.457.4081B}, reinforce that star-forming galaxies can reach X-ray luminosities that overlap the AGN regime, particularly at the low metallicities present in an SDSS-selected sample spanning a range of masses.}

\begin{figure}
    \centering
    \includegraphics[width=\columnwidth]{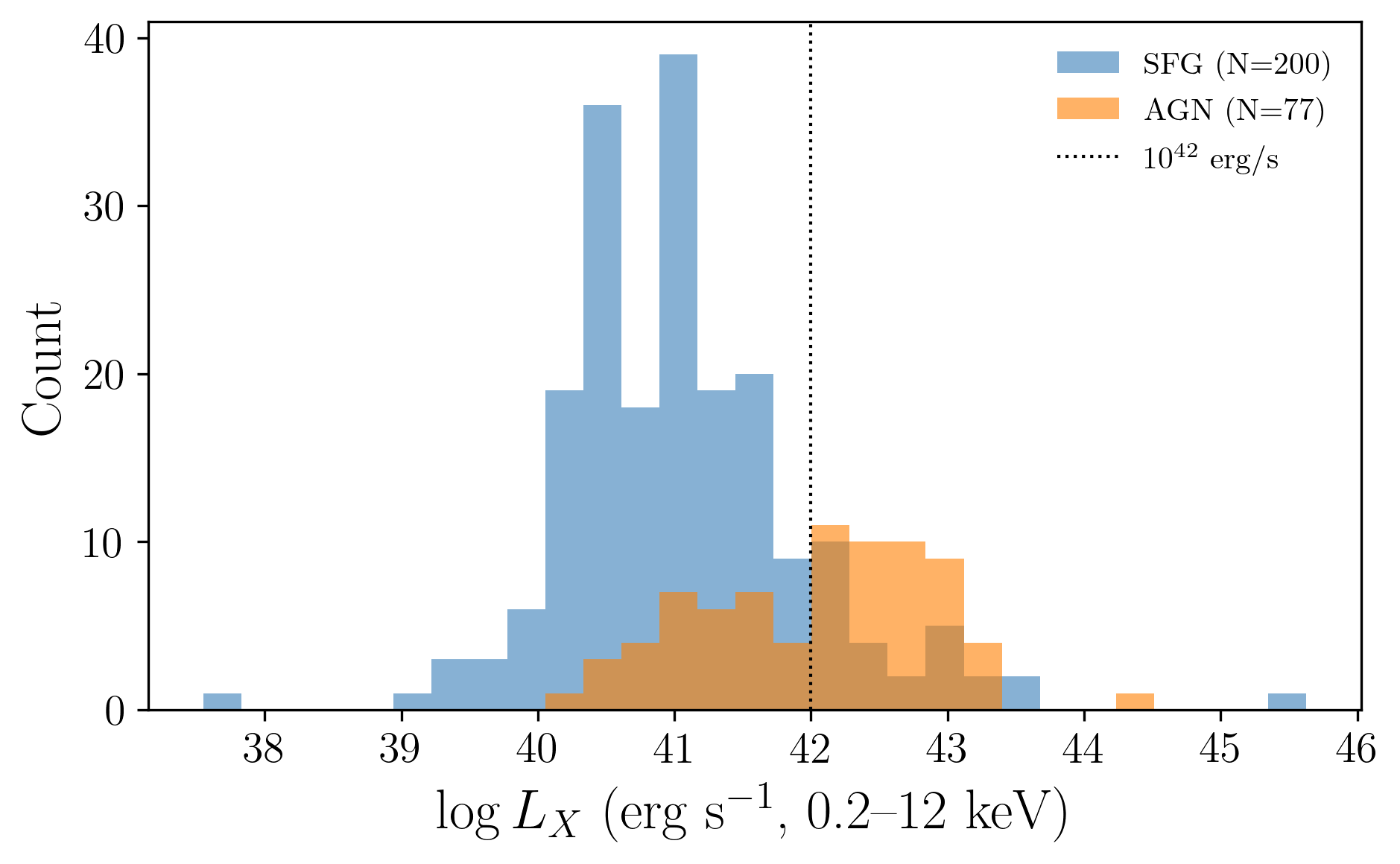}
    \includegraphics[width=\columnwidth]{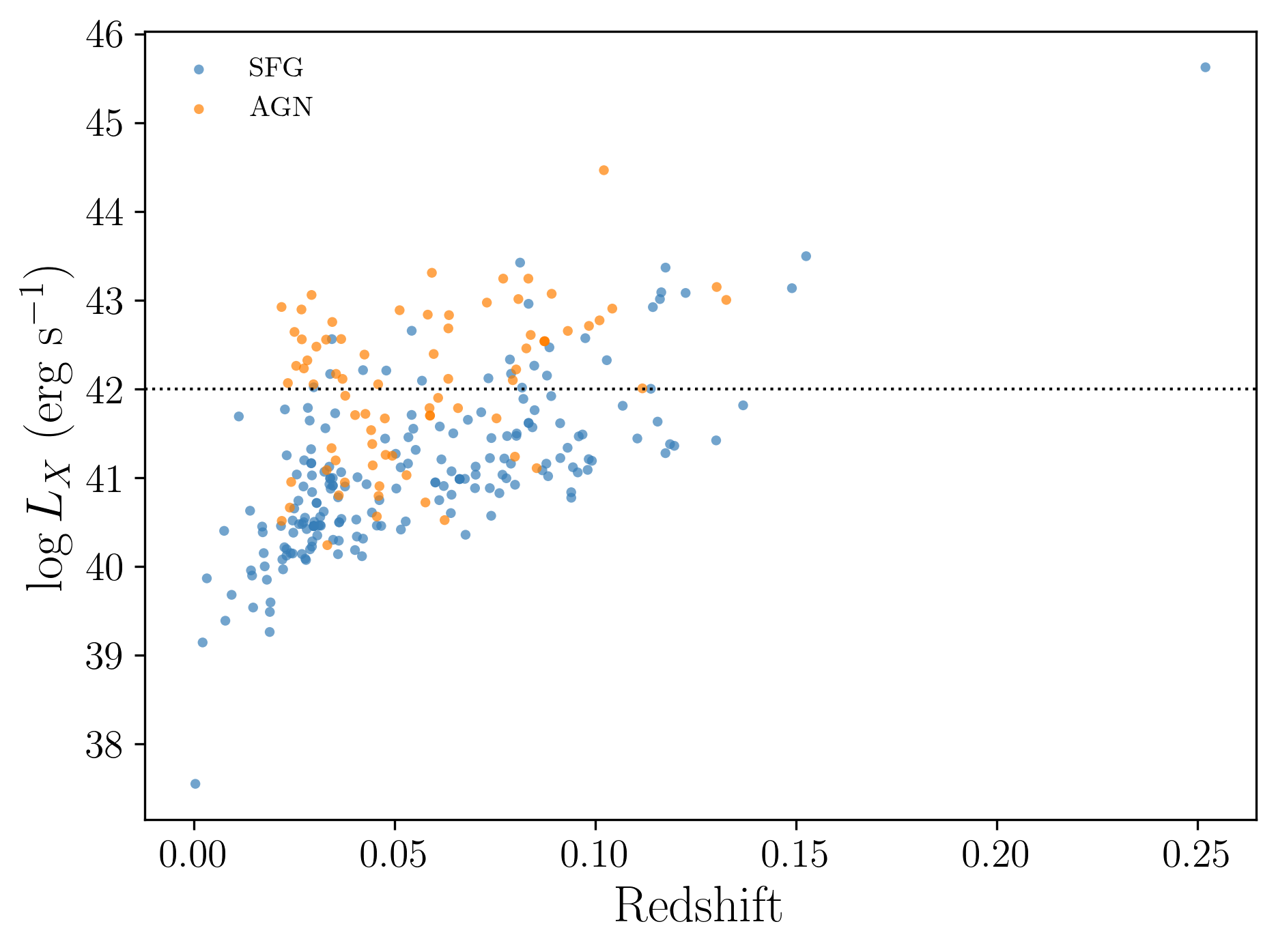}
    \caption{The distribution of the X-ray luminosities of the galaxies in the dataset. The top panel shows the luminosity histogram for both classes, with significant overlap between the two across the $10^{42}$ erg/s threshold. The bottom panel shows luminosity as a function of redshift, illustrating that both classes track the XMM flux limit, where objects at higher redshifts are only detected at higher luminosities, indicating that the X-ray subsample is luminosity-biased and not representative of the broader SDSS population. \textrm{Only objects with a confirmed X-ray detection are shown; non-detections are excluded rather than being included as upper limits or via survival analysis. The 41.6\% of AGN below and 13.5\% of SFGs above the $10^{42}$ erg/s threshold quantify the overlap.}}
    \label{fig:luminositydistribution}
\end{figure}

\textrm{We can quantify this overlap in our dataset directly. Computing observed-frame 0.2--12 keV luminosities for the X-ray subsample, we find that 41.6\% of AGN (32/77) fall below the canonical $10^{42}$ erg/s threshold, while 13.5\% of SFGs (27/200) fall above it. \textrm{We neglect k-corrections, which are small and only weakly class-dependent at these redshifts. For an AGN power law with $\Gamma \approx 1.8$ and a softer SFG spectrum with $\Gamma \approx 2.0-2.5$, the differential correction is 0.004-0.014 dex at the sample median redshift ($z \approx 0.05$) and at most 0.07 dex at $z = 0.25$. Applying these corrections moves one AGN below the threshold (33/77, 42.9\%) and leaves the SFG fraction unchanged, and so the overlap is robust to this choice. These fractions refer only to the X-ray-detected subsample. Because the XMM flux limit excludes fainter sources at higher redshift (as shown in the bottom panel of Figure \ref{fig:luminositydistribution}), they are not population-level estimates, and may be biased by the detection threshold. A survival analysis incorporating non-detections would be required for a population estimate, but the qualitative conclusion that the threshold is not a clean separator holds within the detected sample.} The median luminosities of the two classes ($\log L_X = 42.1$ for AGN and $41.0$ for SFGs) differ by only about an order of magnitude and overlap substantially. The threshold is therefore not a clean separator for this moderate-luminosity, optically-selected sample: a large fraction of AGN are X-ray-faint enough to resemble SFGs, and a non-negligible tail of SFGs reach AGN-like luminosities through HMXB emission.}

This overlap creates an issue with the use of X-ray flux as a classification feature in a machine learning context. The prevailing theory suggests that AGN will generally exhibit higher X-ray luminosities than SFGs. For high-luminosity AGN, this holds true; however, the galaxies in the SDSS sample we used are not, on average, high-luminosity. They are predominantly moderate-luminosity AGN and moderately star-forming galaxies at relatively low redshifts, drawn from a spectroscopic survey that is sensitive to optical emission lines rather than X-ray flux, as shown in Figure \ref{fig:luminositydistribution}. This selection means that the AGN in the sample are likely to cluster at moderate X-ray luminosities, a region where the HMXB contribution from SFGs would be non-negligible. As such, while the prevailing theory may be correct as a statement about X-ray selection in general, it may still be a poor guide to the behavior of X-ray features in this specific sample, as the sample’s AGN luminosity distribution places it squarely in the overlap that \citet{2012MNRAS.419.2095M} describe.

This suggests that the prevailing theory is not so much incorrect as it is incomplete when applied to moderate-luminosity, optically-selected samples of AGN. The theory was developed primarily in the context of deep X-ray surveys optimized to find AGN, where the selection function favors high-luminosity objects and where X-ray coverage is deep enough to detect even relatively faint sources. However, when that theory is imported into a multi-wavelength classification framework based on a sample selected for optical spectroscopic completeness rather than X-ray brightness, its predictions may not fully transfer. Since the SDSS sample in \citet{2024PASP..136l4102D} was not selected to maximize X-ray completeness or to probe high-luminosity AGN, it would be subject to upper limits and non-detections for the fainter sources. In this context, the X-ray flux values available to the model would not be the sharp discriminators that the prevailing theory envisions, but rather shallow, heterogeneous measurements that may contain more noise than signal for the moderate-luminosity objects that dominate the sample.

Therefore, the provisional conclusion on the astrophysical side is nuanced: the prevailing theory is likely correct in the regime for which it was developed, but does not fully extend to the sample we previously used. The overlap between the AGN and SFG X-ray properties in this sample means that the X-ray flux would carry less discriminatory information than the theory would suggest. This is not merely an artifact of the data, but \textrm{a real astrophysical effect: it is part of why the X-ray feature carries little discriminatory signal to begin with, and thus why the model gains little from it.}

\section{Machine Learning Considerations: Label Noise and Feature Sparsity}
\label{sec:ml}

\subsection{Instance-Dependent Label Noise}

\textrm{However, luminosity overlap alone does not fully account for the outcome. Even setting aside the sample-selection effect quantified below, the X-ray feature fails to improve accuracy despite being physically informative about AGN activity. This failure to help, rather than a simple absence of signal, points to an additional artifact within the machine learning process itself.}

In our previous paper, we utilized a Random Forest classifier, an ensemble machine learning model that constructs a large number of decision trees and aggregates their predictions by a majority vote. It is well-suited to multi-wavelength classification problems such as this one, as it handles high-dimensional feature spaces, is robust to individually noisy features, and provides built-in feature importance estimates. However, like all supervised learning methods, it is still fundamentally dependent on the quality and representativeness of its training labels.

The SDSS AGN/SFG classifications we used as training labels were derived from a combination of optical emission-line criteria and human inspection \citep{2012AJ....144..144B}. This means that the emission-line criteria are essentially a formalized version of the BPT diagram; consequently, the training labels are correlated, by construction, with the optical emission-line features that also serve as the model’s primary non-X-ray inputs. The model is being asked to learn a classification rule that its own labels already encode in the optical features, and the X-ray flux is among the only features that carry information independent of the label-generation process. This structural coupling between labels and features is what \citet{2020arXiv200708199S} describe as label noise.

In fact, their framework of instance-dependent label noise is particularly relevant here. In instance-dependent noise, the probability that a label is incorrect is not constant across the dataset but depends on the properties of the individual data point. In the context of SDSS data, this translates to a higher probability of misclassification for galaxies near the BPT demarcation line, or composite objects that share the properties of both AGN and SFGs and are genuinely ambiguous under optical classification criteria. \textrm{This is not merely a statistical artifact of the classifier: \citet{2019ApJ...876...12A} show directly that X-ray-selected AGN can fall within the star-forming region of the BPT diagram, meaning a real population of genuine AGN receives star-forming labels under optical classification. \textrm{Optical labels are therefore known to be imperfect in precisely the way the instance-dependent label noise framework describes.} The optical mislabeling of genuine X-ray AGN is well documented: X-ray-detected but optically weak ``XBONG'' and ``optically dull'' AGN \citep{2003MNRAS.344L..59M,2007A&A...470..557C,2006ApJ...645..115R}, ``lineless'' AGN with low Eddington ratios \citep{2011ApJ...733...60T}, and X-ray AGN falling in the star-forming region of the BPT diagram \citep{2014A&A...568A.108P,2016A&A...594A..72P,2023ApJ...943..174A} all describe populations for which optical line ratios understate nuclear activity. The scatter in the $L_X-L_{[\text{OIII}]}$ relation \citep{2006A&A...455..173P,2010ApJ...722..212T} further limits optical line strength as a proxy for AGN power. This mislabeling is strongly mass-dependent: \citet{2020MNRAS.492.2268B} find that optical diagnostics fail to identify 85\% of X-ray-selected AGN in dwarf galaxies, and \citet{2022MNRAS.510.4556B} show that the agreement between optical and X-ray AGN identification rises monotonically with host stellar mass, from 30\% in the lowest-mass bin to 93\% in the highest. Because our SDSS sample spans a wide range of stellar masses, this mass-dependent optical incompleteness is a direct, quantified source of the instance-dependent label noise our mechanism describes.} These are exactly the objects for which X-ray data would be most useful as an independent discriminator, as all other features would be the least reliable here. However, these are also the objects for which the model’s label noise would be the highest, meaning that, when X-ray features are introduced as a new signal, the model must simultaneously try to learn using an actually informative feature and, at the same time, navigate a label space that is most uncertain precisely where that feature is most useful. Under these conditions, the addition of X-ray features \textrm{would introduce a signal that conflicts with the label structure rather than reinforcing it.}

The model encounters galaxies near the BPT boundary for which the X-ray flux points toward one classification, while the optical emission-line features and the labels those features encode point toward another. A Random Forest trained to minimize classification error on the training labels would, when X-ray features are included, resolve this conflict by down-weighting, or effectively ignoring, the X-ray signal, as incorporating it increases training loss on the noisy near-boundary objects even if it would improve generalization to the true underlying categories. \textrm{The result is that the feature is heavily down-weighted. Since it is informative about the true physical categories, but \textrm{un}informative with respect to the noisy labels, it is driven to low weight by a model that optimizes for the labels rather than the underlying classes. This predicts two testable outcomes\textrm{: adding the feature should leave both overall and class-resolved performance essentially unchanged, and the trained model should make little use of it.} We examine both directly.}

\textrm{To test this prediction and separate the effect of the X-ray feature from the effect of the sample restriction, we evaluate three configurations under 5-fold cross-validation. Of the 312 objects with X-ray detections, 277 have complete photometry across all feature bands and enter the models; the remainder are excluded to avoid the classifier learning from patterns of missingness rather than from physical measurements. The three configurations are: (a) the full optical sample (25{,}668 objects with complete photometry, optical features only); (b) the X-ray subsample (277 objects, optical features only); and (c) the X-ray subsample (277 objects, optical features plus X-ray flux). Configuration (a) achieves an accuracy of $97.37\% \pm 0.12\%$, (b) achieves $93.85\% \pm 2.10\%$, and (c) achieves $93.86\% \pm 0.98\%$, where uncertainties denote the standard deviation across folds. Comparing (a) and (b) isolates the effect of restricting to the X-ray-detected subsample, a decrease of 3.52 percentage points driven purely by the change in sample; comparing (b) and (c) isolates the effect of adding the X-ray feature on a fixed sample, a change of -0.01 percentage points. The near-identical accuracy of (b) and (c) indicates that the X-ray feature has negligible effect on overall accuracy once the sample is held fixed: the Random Forest makes little use of it. As we show below, \textrm{this holds at the class level and in the feature's permutation importance}. The apparent accuracy decrease reported in \citet{2024PASP..136l4102D} therefore reflects predominantly the harder, boundary-concentrated X-ray subsample rather than the X-ray feature itself.}

\textrm{Because the X-ray subsample is imbalanced ($\sim$2.6:1), aggregate accuracy could hide changes in minority-class performance. We therefore also report class-resolved metrics for configurations (b) and (c) in Table \ref{tab:classresolvedmetrics}. Adding the X-ray feature leaves the pooled out-of-fold confusion matrix unchanged: 68 of 77 AGN and 192 of 200 SFGs are correctly classified in both configurations, giving identical AGN precision (0.895), AGN recall (0.883), and AGN F1 (0.889). The two configurations disagree on only four objects, split evenly between them (exact McNemar $p = 1.00$, both overall and within each class). ROC-AUC (0.987 vs. 0.986) and PR-AUC for the AGN class (0.967 vs. 0.964) are likewise indistinguishable. The negligible effect of the X-ray feature therefore holds at the class level and is not an artifact of aggregate accuracy.}

\begin{table}
    \centering
    \caption{\textrm{Class-resolved performance of configurations (b) and (c) on the 277-object X-ray subsample (77 AGN, 200 SFGs). Values quoted in the text are computed from the pooled predictions.}}
    \label{tab:classresolvedmetrics}
    
    \begin{tabular}{lcc}
        \hline
        Metric & (b) Optical only & (c) Optical + X-ray \\
        \hline
        Accuracy      & $0.939 \pm 0.021$ & $0.939 \pm 0.010$ \\
        AGN precision & $0.897 \pm 0.052$ & $0.898 \pm 0.046$ \\
        AGN recall    & $0.882 \pm 0.058$ & $0.883 \pm 0.031$ \\
        AGN F1        & $0.888 \pm 0.040$ & $0.889 \pm 0.015$ \\
        SFG precision & $0.956 \pm 0.020$ & $0.955 \pm 0.010$ \\
        SFG recall    & $0.960 \pm 0.022$ & $0.960 \pm 0.022$ \\
        SFG F1        & $0.958 \pm 0.014$ & $0.958 \pm 0.008$ \\
        ROC-AUC       & $0.989 \pm 0.005$ & $0.987 \pm 0.008$ \\
        PR-AUC (AGN)  & $0.974 \pm 0.011$ & $0.969 \pm 0.019$ \\
        \hline
        \multicolumn{3}{l}{\textit{Pooled confusion matrix (true $\rightarrow$ predicted)}} \\
        AGN $\rightarrow$ AGN & 68  & 68  \\
        AGN $\rightarrow$ SFG & 9   & 9   \\
        SFG $\rightarrow$ AGN & 8   & 8   \\
        SFG $\rightarrow$ SFG & 192 & 192 \\
        \hline
    \end{tabular}
\end{table}

\begin{figure}
    \centering
    \includegraphics[width=\columnwidth]{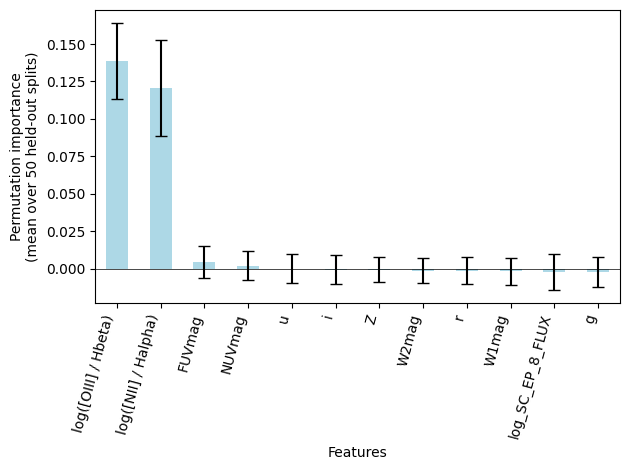}
    \caption{A permutation importance plot of the features, \textrm{averaged over 50 held-out splits (5-fold cross-validation repeated ten times with 100 permutations per split). The error bars show the standard deviation across splits. As is shown in the figure, the X-ray flux feature ($\texttt{log\_SC\_EP\_8\_FLUX}$) has permutation importance consistent with being zero.}}
    \label{fig:permutationimportance}
\end{figure}

\textrm{We estimate the permutation importance of the X-ray feature across repeated held-out splits rather than a single test set, since on a 277-object sample, a single split captures permutation variance but not data-resampling variance. Over the five cross-validation folds, the mean importance is $-0.0009 \pm 0.0094$ in accuracy; over 50 held-out splits (5-fold cross-validation repeated ten times), it is $-0.0021 \pm 0.0120$ in accuracy and $0.0012 \pm 0.0163$ in balanced accuracy, with negative and positive values about equally frequent, as shown in Figure \ref{fig:permutationimportance}. The X-ray feature's permutation importance is therefore consistent with being zero. We note that a single-split estimate on this sample can appear significantly non-zero, showing that permutation importance on small samples should be resampled across splits before being interpreted. Together with the decomposition and class-resolved metrics, this indicates that the trained Random Forest makes essentially no use of the X-ray feature.}

A further issue within the machine learning process is sample size and feature sparsity. The XMM-Newton Survey, while among the most sensitive X-ray observatories available, covers only a fraction of the sky through targeted pointings rather than an all-sky survey, meaning that only SDSS objects coincidentally within existing XMM fields received X-ray coverage. For the moderate-luminosity objects in the SDSS sample, a significant fraction will have no X-ray detections at all. \textrm{Including such objects would require substituting upper limits or otherwise imputing values, producing a feature that mixed genuine detections with non-physical fill-ins and statistical noise, which we avoided by restricting the X-ray model to objects with genuine detections.} When a feature is both sparse in coverage and uncertain in measurement, a Random Forest model will typically assign it low feature importance, not because the feature is intrinsically uninformative, but because the measured values prove unreliable for making predictions. While our original X-ray sub-sample was more balanced (\textasciitilde{}2:1) than the parent population (\textasciitilde{}7:1), the extreme sparsity of X-ray detections (only 312 objects\textrm{, of which 277 have complete photometry and enter the models}) limited the Random Forest's ability to find a robust decision boundary. \textrm{This sparsity is not remediable by data synthesis. Because XMM-Newton observed only the fraction of the sky falling within its targeted pointings, no X-ray measurement exists for the majority of objects in the full optical sample. Imputing or bootstrapping X-ray values for these objects would not recover the missing information: any synthesized value would either encode the non-random pattern of which objects were observed (which correlates with source properties and would leak class information) or fabricate a flux with no physical basis. The X-ray feature is therefore necessarily restricted to the subsample of genuinely detected objects, and the comparison between the full-sample and X-ray-subsample models cannot be equalized by imputation.} In this low-data regime, even a moderate class imbalance can exacerbate the impact of label noise. This reduces sensitivity to AGN-discriminating features, including X-ray flux, and \textrm{contributes to the feature being assigned \textrm{negligible} importance by the ensemble.}

\subsection{Boundary-Concentration of Misclassifications}

\begin{figure}
    \centering
    \includegraphics[width=\columnwidth]{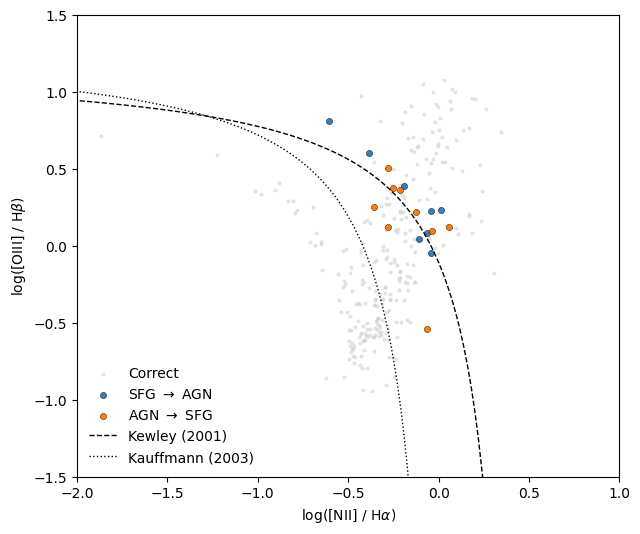}
    \caption{\textrm{BPT diagram of the 277-object X-ray subsample, with class predictions aggregated across all five cross-validation folds. Of the 277 objects, 17 are misclassified: blue points are SFGs misclassified as AGN and orange points are AGN misclassified as SFGs, while grey points are correctly classified. The dashed and dotted curves show the \citet{2001ApJ...556..121K} and \citet{2003MNRAS.346.1055K} demarcation lines respectively.}}
    \label{fig:xraysub}
\end{figure}

\begin{figure}
    \centering
    \includegraphics[width=\columnwidth]{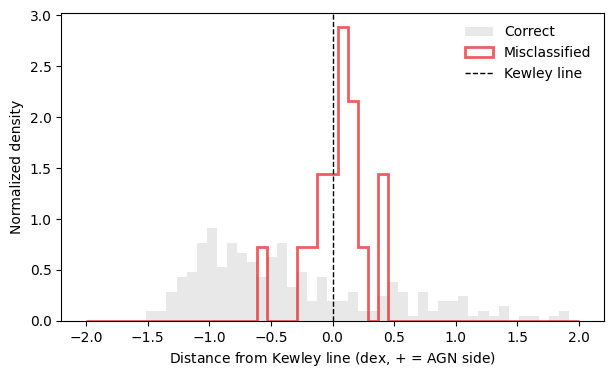}
    \caption{\textrm{Normalized distribution of signed distances from the \citet{2001ApJ...556..121K} demarcation line for correctly classified ($N=260$) and misclassified ($N=17$) objects in the 277-object X-ray subsample, evaluated via 5-fold cross-validation. Distance is the vertical offset in $\log\left([\text{OIII}]/\text{H}\beta\right)$ at fixed $\log\left([\text{NII}]/\text{H}\alpha\right)$, positive on the AGN side. Misclassified objects concentrate near zero while correctly classified objects span a broad range.}}
    \label{fig:xrayhist}
\end{figure}

To directly test whether the \textrm{misclassifications are} boundary-concentrated, as the instance-dependent label noise framework predicts, we examined the BPT positions of objects misclassified by the X-ray model using 5-fold cross-validation across all 277 objects with X-ray detections and complete photometry. As shown in Figures \ref{fig:xraysub} and \ref{fig:xrayhist}, misclassified objects cluster sharply near the Kewley line, with a median distance of 0.123 dex from the boundary, compared to 0.743 dex for correctly classified objects. To quantify this boundary-concentration, we apply a two-sided Kolmogorov-Smirnov test, which measures the maximum difference between the cumulative distance distributions of the two groups without assuming a particular distributional form. The test yielded $D = 0.745,\; p = 1.67\times 10^{-9}$, indicating that the boundary-concentration of misclassified objects relative to correctly classified ones is statistically significant well beyond chance, and that the two populations are drawn from fundamentally different distance distributions. \textrm{A one-sided Mann-Whitney $U$ test further establishes the direction of this difference: correctly classified objects lie significantly farther from the boundary than misclassified ones ($U=4002,\;p=1.08\times 10^{-8}$).} Figure \ref{fig:xrayhist} shows that the distribution of misclassified objects spikes on both sides of the boundary near zero, while correctly classified objects are distributed broadly across the parameter space. This boundary-concentration is symmetric across both misclassification directions, indicating that the effect is driven by proximity to the boundary rather than systematic bias for either class. \textrm{It is consistent with the expected signature of instance-dependent label noise but is not unique to it. Objects near the Kewley line may also be physically composite systems in which star formation and nuclear activity both contribute to the line emission, so that a binary label is ambiguous even when assigned in good faith. Our diagnostic cannot distinguish an individual mislabeled AGN from a genuinely intermediate system: the literature above establishes that optical mislabeling of X-ray AGN occurs, but not that it accounts for these particular objects. We note further that 15 of the 17 misclassified objects are misclassified both with and without the X-ray feature, so the X-ray flux does not resolve these boundary cases. Whether this is because the X-ray signal conflicts with labels that are noisy at the boundary or because it is redundant against labels already determined by the optical features cannot be separated using BPT-derived labels alone. Doing so would require labels independent of the optical line ratios (as is discussed in Section \ref{sec:conclusion}).}

\textrm{Thus, the evidence suggests that the X-ray feature's failure to improve performance and the apparent decrease reported previously is most plausibly explained by the interaction of sample selection, instance-dependent label noise, X-ray feature sparsity, and class imbalance, all of which act to suppress the model's ability to utilize the discriminatory information that the X-ray data does actually contain. We would like to note that we measure boundary distance from the Kewley line for consistency with our labels; the Kauffmann line and its recent modifications \citep{2019ARA&A..57..511K,2021ApJ...922..156A} may better trace the star-forming locus; nonetheless, our choice of demarcation does not affect the qualitative boundary-concentration result.}

\section{Discussion}

Having examined both the astrophysical and machine learning explanations separately, we now assess their relative contributions to the observed discrepancy. It seems most likely that the discrepancy is primarily a consequence of sample selection and is compounded by a genuine, but secondary, astrophysical overlap. \textrm{The controlled decomposition in Section~\ref{sec:ml} shows that the apparent 8.25\% accuracy drop reported in \citet{2024PASP..136l4102D} is driven predominantly by sample selection: restricting to the X-ray-detected subsample lowers accuracy by 3.52 percentage points before any X-ray feature is introduced, while the feature itself changes accuracy by a negligible -0.01 percentage points on a fixed sample. \textrm{The class-resolved metrics and resampled permutation importance confirm that this null effect is not an artifact of aggregate accuracy: the confusion matrix is unchanged and the feature's importance is consistent with zero. The relevant question is thus not why X-rays reduce accuracy, but why a feature that is physically informative about AGN activity goes essentially unused once the sample is fixed. Our diagnostics point to the label structure. Since the training labels are largely determined by the optical features, misclassifications concentrate at the boundary where those labels are the least reliable, as is shown by the fact that the median boundary distance of misclassified objects (0.123 dex) is approximately six times smaller than that of correctly classified objects (0.743 dex), and the X-ray feature does not resolve these boundary cases. Whether it goes unused because it conflicts with noisy labels at the boundary or because it is redundant against labels already encoded in the optical features cannot be separated with BPT-derived labels, but in either case, a label-optimizing model has no incentive to use a signal independent of how its labels were generated.}}

\section{Conclusion}
\label{sec:conclusion}

Ultimately, the \textrm{apparent} discrepancy found in our previous paper is best understood not as a failure of the theory or the model, but rather as a diagnostic revealing the tension between two bodies of knowledge that have developed largely in parallel. Resolving that tension will require work in both high-energy astrophysics and machine learning: more careful accounting of HMXB contamination and luminosity-dependent selection effects on the astrophysical side; and cleaner labels, deeper X-ray coverage, and noise-robust training methods on the machine learning side. \textrm{One concrete route to independent labels is the $L_X$-SFR excess selection of \citet{2019ApJ...876...12A}, which identifies X-ray AGN without reference to optical line ratios. Applying it here would require star-formation-rate estimates not available for most of our X-ray-detected sample, but represents a natural next step. \textrm{In addition, our X-ray feature is a single full-band (0.2-12 keV) flux. Because the AGN/HMXB distinction also depends on spectral shape, a hard-band (2-10 keV) flux or hardness ratio would reduce the contribution of soft thermal emission from SFGs and would be a natural next feature to test.} Alternative classification schemes that reduce reliance on the full BPT line set, such as the WHaD diagram of \citet{2024A&A...682A..71S}, which classifies ionizing sources from the H$\alpha$ equivalent width and velocity dispersion alone and remains usable when the weaker BPT lines fall below the required signal-to-noise, offer another route to labels less coupled to the specific optical features driving the noise. Testing whether the X-ray feature's discriminatory power recovers under such independent labels is a natural extension of this work.} Further development along both lines of work will be necessary to definitively answer the question of whether X-ray data serve as a reliable distinction between AGN and SFGs in optically-selected samples.

There are two primary implications of this conclusion. First, for classifying galaxies, the most immediate takeaway is that future classifiers should look to either use independent AGN selection criteria to construct their training labels or apply training that is label noise-robust. Second, there exists a substantive limitation in the prevailing theory of X-ray AGN diagnostics when applied to the majority of galaxies in large spectroscopic surveys like SDSS. As these surveys are used increasingly as training grounds for machine learning classifiers intended to operate at scale across future surveys, the domain mismatch between the theory’s native regime and the survey’s actual population will become a more pressing concern.

It is worth noting that this conclusion is neither definitive nor exhaustive. It is the best-supported conclusion given the available evidence, and not a claim that the astrophysical question is settled. In a hypothetical future model trained on independent, non-BPT-derived labels, one would still expect some degradation in X-ray discriminatory power for the moderate-luminosity SDSS objects, because the overlap between HMXB-bright SFGs and low-luminosity AGN would still be present in the data. The \textrm{interaction of sample selection and label noise} amplifies and makes visible what is otherwise a moderate astrophysical limitation. Removing \textrm{this interaction} would not make X-ray classification trivial, but would allow for a more honest assessment of how much discriminatory power X-ray data actually carries in this regime.

\section*{Acknowledgements}

I would like to thank Dr. Antonio Rodriguez for his mentorship and work on our 2024 paper, and my professor, Dr. Allen Durgin, for his editorial guidance and advice.

%%%%%%%%%%%%%%%%%%%%%%%%%%%%%%%%%%%%%%%%%%%%%%%%%%
\section*{Data Availability}

The data underlying this study are publicly available from the following astronomical survey archives: the Sloan Digital Sky Survey (SDSS) via the \href{https://www.sdss.org/dr19/data_access}{SDSS data portal}; the Wide-field Infrared Survey Explorer (WISE) via the \href{https://irsa.ipac.caltech.edu/frontpage/}{NASA/IPAC Infrared Science Archive (IRSA)}; the Galaxy Evolution Explorer (GALEX) via the \href{https://archive.stsci.edu/missions-and-data/galex}{Mikulski Archive for Space Telescopes (MAST) GALEX archive}; and X-ray data from the XMM-Newton Serendipitous Source Catalogue via the \href{https://heasarc.gsfc.nasa.gov/w3browse/xmm-newton/xmmssc.html}{XMM-Newton Science Archive / HEASARC XMMSSC archive}. \textrm{The code and data used in this paper are available upon reasonable request to the author.}

\section*{Conflict of Interest}

\textrm{There are no conflicts of interest to declare for this paper.}



\bibliographystyle{rasti}
\bibliography{example} % if your bibtex file is called example.bib

%\begin{thebibliography}{99}
%\end{thebibliography}

%%%%%%%%%%%%%%%%%%%%%%%%%%%%%%%%%%%%%%%%%%%%%%%%%%

%%%%%%%%%%%%%%%%% APPENDICES %%%%%%%%%%%%%%%%%%%%%

% \appendix

% \section{Some extra material}

% If you want to present additional material which would interrupt the flow of the main paper,
% it can be placed in an Appendix which appears after the list of references.

%%%%%%%%%%%%%%%%%%%%%%%%%%%%%%%%%%%%%%%%%%%%%%%%%%

% Don't change these lines
\bsp	% typesetting comment
\label{lastpage}
\end{document}